\documentclass[10pt]{iopart}

\usepackage{iopams} 
\usepackage{setstack}
\usepackage{bm}
\usepackage{color}
\usepackage{xcolor}
\usepackage{graphicx}
\usepackage{chemarrow}
\usepackage{multirow}
\usepackage{slashbox}
\usepackage{cite}
\usepackage{theorem}
\newtheorem{defin}{Definition}
\newtheorem{property}{Property}
{\theorembodyfont{\upshape}
\newtheorem{assump}{Assumption}
\newtheorem{remark}{Remark}
\newtheorem{example}{Example}

\newtheorem{lemma}{Lemma}
}
\begin{document}

%\title[\SMS]{Identification and nonlinearity compensation of hysteresis using NARX models}

\title[Identification and nonlinearity compensation of hysteresis]{Identification and nonlinearity compensation of hysteresis using NARX models}

\author{Petrus E. O. G. B. Abreu$^1$, Lucas A. Tavares$^1$, Bruno O. S. Teixeira$^2$ and Luis A. Aguirre$^2$}

\address{$^1$ Graduate Program in Electrical Engineering, Universidade Federal de Minas Gerais, Av. Ant\^onio Carlos 6627, 31270-901, Belo Horizonte, MG, Brazil}
\address{$^2$ Departament of Electronic Engineering, Universidade Federal de Minas Gerais, Av. Ant\^onio Carlos 6627, 31270-901, Belo Horizonte, MG, Brazil}
\eads{\mailto{petrusabreu@ufmg.br}, \mailto{amarallucas@ufmg.br}, \mailto{brunoot@ufmg.br} and \mailto{aguirre@ufmg.br}}
\vspace{10pt}
\begin{indented}
\item[]December  2019
\end{indented}

\begin{abstract}
This paper deals with two problems: the identification and compensation of hysteresis nonlinearity in dynamical
systems using nonlinear polynomial autoregressive models with exogenous inputs (NARX). First, based on gray-box
identification techniques, some constraints on the structure and parameters of NARX models are proposed to ensure
that the identified models display a key-feature of hysteresis. In addition, a more general framework is developed
to explain how hysteresis occurs in such models. Second, two strategies to design hysteresis compensators
are presented. In one strategy the compensation law is obtained through simple algebraic manipulations performed
on the identified models.  It has been found that the compensators based on gray-box models
outperform the cases with models identified using black-box techniques.
In the second strategy, the compensation law is directly identified from the data.
Both numerical and experimental results are presented to illustrate
the efficiency of the proposed procedures.
\end{abstract}

\vspace{2pc}
\noindent{\it Keywords}: Hysteresis, gray-box identification, compensation of nonlinearities, NARX model
%\submitto{\SMS}
%\maketitle

%==================================================================
\section{Introduction}

Hysteresis is a nonlinear behavior that is present in several systems and devices.
It is commonly related to the phenomena of ferromagnetism, plasticity, and friction,
among others \cite{Visintin1994}. Some examples include mechanical, electronic and
biomedical systems, as well as sensors and actuators such as magneto-rheological
dampers, piezoelectric actuators and pneumatic control valves
\cite{Choudhury_etal2008,Rakotondrabe2013,Peng_Chen2013}. An intrinsic feature of
such systems is the memory effect, meaning that their output depends on the history
of the corresponding input.

In addition to the memory effect, the literature provides different definitions
and conditions to distinguish such systems and characterize the hysteretic behavior.
In some cases, the occurrence of hysteresis has been associated with the existence
of several equilibrium points whenever these systems are subject to a constant
\cite{Morris2011} or time-varying \cite{Martins_Aguirre2016} input signal.
Additionally, hysteresis has also been defined as a hard nonlinearity that depends on
the magnitude and rate of the input signal. These aspects can pose various performance
limitations if not properly taken into account during the control design
\cite{Tao_Kokotovic1995,Rakotondrabe2013}. Hence, a common goal is to attenuate the
hysteretic behavior of the system \cite{Visone2008,Chaoui_Gualous2016,Yi_etal2019}.

In many approaches, the compensation of hysteresis starts with obtaining a suitable model.
In the literature, several hysteresis models have been proposed based on phenomenological,
black-box and gray-box modeling approaches.

In the realm of models based on first principles, important contributions have been made
based on differential equations and operators \cite{Hassani_etal2014}, such as the Bouc-Wen
model \cite{Wen1976}, the Duhem model \cite{JinHyoungOh_Bernstein2005}, the Preisach model
\cite{Ge_Jouaneh1996} and the Prandtl-Ishlinskii operator \cite{Brokate_Sprekels1996}.
These models have been widely used to predict the hysteresis behavior due to their ability to
describe a variety of hysteresis loops that resemble the proprieties of a wide class of real
nonlinear hysteretic systems \cite{Smyth_etal2002}. Besides, such models are known to be challenging
for system identification techniques \cite{Quaranta_etal2020}. Apart from the computational effort required
in the identification of phenomenological models, their application in the design of compensators
is somewhat limited due to their structural complexity \cite{Peng_Chen2013,Hassani_etal2014}.

Black-box modeling does not rely on prior knowledge about the system
\cite{Chan_etal2015,Ayala_etal2015,Fu_etal2016}. Unfortunately, relevant features that should be
present in a model to reproduce hysteresis and an appropriate structure for designing compensators
are not ensured by black-box techniques. Hence, the search for models that have specific features,
that are accurate and that have a suitable structure for designing compensators remains an open problem.

Models obtained using gray-box techniques can be tailored to reproduce specific relevant features \cite{Aguirre_2019}.
In this context, nonlinear autoregressive with exogenous inputs (NARX) models are considered a convenient
choice due to their ability to predict a wide class of nonlinear behaviors
\cite{Leontaritis_Billings1985PartI,Leontaritis_Billings1985PartII}. Another interesting feature is the
structural flexibility of such models. Therefore, enforcing constraints on the model structure
(e.g., in order to make it suitable for designing compensators) does not drastically affect its predictive
ability \cite{Pearson1999}. Despite these promising aspects, the literature on this approach for modeling
and compensating of hysteretic systems is scarce
\cite{Leva_Piroddi2002,Deng_Tan2009,Worden_Barthorpe2012,Dong_Tan2014,Martins_Aguirre2016,L_Junior_etal2019}.

In this sense, an important step in modeling the hysteresis nonlinearity was advanced in
\cite{Martins_Aguirre2016}, in which sufficient conditions are presented for NARX models
to display a hysteresis loop when subject to a certain class of input signals. The concept
of a bounding structure ${\cal H}$ formed by sets of stable equilibria and its implication
on the existence of the hysteresis loop in the identified models has also introduced in \cite{Martins_Aguirre2016}.
However, for more general cases, this concept and conditions need to be adapted.
For instance, the conditions proposed in \cite{Martins_Aguirre2016} are not sufficient to
ensure the existence of several equilibrium points at steady-state \cite{Bernstein2007,Morris2011}.
Also, the concept of bounding structure is limited to cases in which the sets of equilibria that
form this structure are stable. Recently, a NARX model was identified for an experimental electronic
circuit with hysteresis \cite{L_Junior_etal2019}. In this paper, we will propose ways to overcome
some shortcomings pointed out in the aforementioned references and a more general framework will
be put forward to explain how hysteresis takes place in identified models.

The main contributions of this work are: the proposition of a specific parameter constraint that ensures reproducing a
key-feature of hysteresis through identified NARX models. A framework is put forward to explain
how the hysteresis loop results from an interplay of attracting and repelling
regions in the input-output plane. Moreover, some structural specifications are enforced during the identification
procedure in such a way that the identified NARX model can be effectively used to mitigate the
hysteresis nonlinearity. Hence, two model-based compensation strategies are introduced.
In the first the compensation law is obtained through simple algebraic manipulations performed
on the identified models.  It has been found that the compensators based on gray-box models
outperform the cases with models identified using black-box techniques.
In the second strategy, the compensation law is directly identified from the data.

%One of the aims of this work is to contribute to a more general understanding of the behavior of
%identified models for hysteretic systems. In particular, a parameter constraint that ensures a
%key-feature of hysteresis is proposed and justified. Also, a framework is put forward to explain
%how, for some models, the hysteresis loop results from an interplay of attracting and repelling
%regions in the $u \times y$ plane. A second aim is to present two model-based compensation strategies.
%It will be seen the usefulness of enforcing some structural {NARXs during the identification
%procedure in such a way that the identified NARX model can be effectively used to mitigate the
%hysteresis nonlinearity. 

This work is organized as follows. Section\,\ref{back} presents the background. A constraint
to ensure hysteresis in the identified models and a framework for understanding how the
hysteresis loop is formed are provided in section\,\ref{ish}. Based on NARX models, two
strategies to design compensators are detailed in section\,\ref{Section:Methodology}.
The numerical and experimental results for the model identification and the compensator
design are, respectively, given in sections \ref{Section:NumericalResults} and
\ref{Section:ExperimentalResults}. Section\,\ref{Section:Conclusions} presents the
concluding remarks.

%==================================================================
\section{Background}
\label{back}

A NARX model can be represented as \cite{Leontaritis_Billings1985PartI}:
\begin{equation}
\label{Eq:NARX_Model}
y_k = \tilde{F}\big(y_{k-1},\cdots,y_{k-n_{y}}, \,u_{k-\tau_{\rm d}},\cdots,u_{k-n_u}\big), 
\end{equation}

\noindent
where $y_k \in \mathbb{R}$ is the output at instant $k \in \mathbb{N}$, $u_k \in \mathbb{R}$ is the input,
$n_y$ and $n_u$ are the maximum lags for the output and input, respectively, $\tau_{\rm d} \in \mathbb{N}^{+}$
is the pure time delay, and $\tilde{F}(\cdot)$ is a nonlinear function
of the lagged inputs and outputs.

This work considers a linear-in-the-parameters extended model set \cite{Billings_Chen1989}
of the NARX model (\ref{Eq:NARX_Model}) with the addition of specific functions, such as
absolute value, trigonometric, and sign function. The goal is to choose functions that
allow the models to predict systems whose nonlinearities cannot be well approximated using
only regressors based on monomials of lagged input and output values. For instance,
\cite{Billings_Chen1989} recommends the addition of absolute value and sine functions as
candidate regressors for the identification of a damped and forced nonlinear oscillator.
In the case of the identification of systems with hysteresis, \cite{Martins_Aguirre2016}
shows that including the regressor given by sign of the first difference of the input, i.e.
${\rm sign}(u_k-u_{k-1})$, in addition to polynomial terms is a sufficient condition to
reproduce hysteresis. Therefore, in this work the models are of the general type:
\begin{eqnarray}
\label{090919a}
y_k & = & F^{\ell}\big(y_{k-1},\cdots,y_{k-n_{y}}, \,u_{k-1},\cdots,u_{k-n_u}, \nonumber \\
& ~ &\hspace{0.4cm} \phi_{1,\,k-1}, \,\phi_{2,\,k-1} \big), 
\end{eqnarray}

\noindent
where $\phi_{1,\,k}{=}u_k-u_{k-1}$, $\phi_{2,\,k} {=} {\rm sign}(\phi_{1,\,k})$, and $F^{\ell}(\cdot)$
is a polynomial function of the regressor variables up to degree $\ell \in \mathbb{N}^{+}$.

Evaluating model (\ref{090919a}) along a data set of length $N$, the resulting set of
equations can be expressed in matrix form as:
\begin{equation}
\bm{y}=\Psi\hat{\bm{\theta}}+\bm{\xi},
\end{equation}

\noindent
where $\bm{y} \triangleq [y_k\,\, y_{k-1}\,\cdots\, y_{k+1-N}]^T \in \mathbb{R}^N$ is the vector of output measurements,
$\Psi \triangleq [\psi_{k-1}^T;\, \cdots;\, \psi^T_{k-N}] \in \mathbb{R}^{N\times n_{\theta}}$ is the matrix composed by
measurements of the regressors vector $\psi_{k-1} \in \mathbb{R}^{n_{\theta}}$ which contains linear and nonlinear
combinations of the variables that compose $F^{\ell}(\cdot)$ in (\ref{090919a}) weighted by the parameter vector
$\hat{\bm{\theta}} \in \mathbb{R}^{n_{\theta}}$, $\bm{\xi} \triangleq[\xi_k\,\, \xi_{k-1}\,\cdots\, \xi_{k+1-N}]^T \in \mathbb{R}^{N}$
is the residual vector and $T$ indicates the transpose.

The unconstrained least squares batch estimator is given by
\begin{equation*}
\hat{\bm{\theta}}_{\rm LS}=(\Psi^T\Psi)^{-1}\Psi^T\bm{y}.
\end{equation*}

\noindent
Assume the set of equality constraints on the parameter vector written as $\bm{c}=S\bm{\theta}$,
where $\bm{c} \in \mathbb{R}^{n_c}$ and $S \in \mathbb{R}^{n_c\times n_{\theta}}$ are known constants.
Then, the constrained least squares estimation problem is given by
\begin{equation}
\label{Eq:ContrainedParameters_Estimator_Problem}
\hat{\bm{\theta}}_{\rm CLS} = \underset{\bm{\theta}\,:\,\bm{c}=S\bm{\theta}}{\arg\min} \big[\bm{\xi}^T\bm{\xi}\big],
\end{equation}

\noindent
whose solution is \cite{Draper_Smith1998}:
\begin{equation}
\hat{\bm{\theta}}_{\rm CLS} {=} \hat{\bm{\theta}}_{\rm LS}{-}(\Psi^T\Psi)^{-1}S^T[S(\Psi^T\Psi)^{-1}S^T]^{-1}(S\hat{\bm{\theta}}_{\rm LS}{-}\bm{c}). \label{Eq:ConstrainedParameters_Estimator}
\end{equation}

For the model structure selection, we use the error reduction ratio (ERR) \cite{Chen_etal1989}
together with Akaike's information criterion (AIC) \cite{Akaike1974}. Other approaches that have
proved to be useful in more demanding contexts are found in
\cite{Piroddi2008,Martins_etal2013,Falsone_etal2015,Retes_Aguirre2019,Araujo_etal2019}.

%==================================================================
\section{Identification of Systems with Hysteresis} 
\label{ish}

The scientific community has been investigating which relevant features must be
present in a model to reproduce hysteresis. Some of these features are:
a characteristic loop behavior displayed on the input-output plane \cite{Bernstein2007},
several stable equilibrium points \cite{Morris2011}, and multi-valued mapping \cite{Deng_Tan2009}.
However, which and how these features can be used in the identification procedure remains an open problem.

Here, a constraint is proposed to ensure a key-feature of hysteresis. Also,
it is shown how the hysteresis loop can be seen as an interplay of attracting
and repelling regions in certain models. Then, the resulting models will be used
to design compensators.

First, a property of hysteresis based only on the main aspect discussed in \cite{Martins_Aguirre2016,Bernstein2007,Morris2011}
is presented. In the sequel, this property is used to obtain constraints on the structure and parameters of the model.

%-------------- Property 1
\begin{property}
	\label{D1}
	An identified model of hysteresis, under a constant input, has two or more real non-diverging equilibria. $\square$
\end{property}
%--------------

In \cite{Martins_Aguirre2016}, Property~\ref{D1} was attained by ensuring that the model had at least
one equilibrium point under loading-unloading inputs, with different values for loading and unloading.
Thus, in (\ref{090919a}) $\phi_{1,\,k}{=}u_k-u_{k-1}$ and $\phi_{2,\,k} {=} {\rm sign}(\phi_{1,\,k})$, with
$\phi_{2,\,k} {=}1$ for loading, and $\phi_{2,\,k} {=}-1$ for unloading.

Hence, hysteresis is a nonlinear behavior that appears in both the static response and the dynamics.
In some works, this nonlinearity is classified as quasi-static because the analyses are performed when
the system is excited by a periodic signal that is very slow compared to the system dynamics
\cite{Ikhouane_Rodellar2007}.

Based on a static analysis of NARX models (\ref{090919a}), we will show which constraints need to
be considered in the identification procedure in order for Property~\ref{D1} to be satisfied. Thereafter,
a quasi-static analysis will be used to describe how hysteresis happens in these models and an illustrative
example will be presented.

%-----------------------------------------------------------------
\subsection{Static analysis}
\label{sa}

By means of static analysis it is possible to determine the fixed points of a model.

%-------------- Assumption 1
\begin{assump}
	\label{R2}
	In order to comply with Property~\ref{D1}, considering the recommendation of the literature,
	the identified models should not have the following regressors:
	\begin{itemize}
	\item[(i)] $y^{p}_{k-\tau_y}$,  $y^{p}_{k-\tau_y}\phi_{1,\,k-\tau_u}^m$
	and $y^{p}_{k-\tau_y}\phi_{2,\,k-\tau_u}^m$
	for $p{>}1,~\forall m$ \cite{Aguirre_Mendes1996},
	\item[(ii)] ${\rm sign}(u_{k-\tau_u}-u_{k-\tau_u-1})^m=\phi^m_{2,\,k-\tau_u}$ for $m>1$ \cite{Martins_Aguirre2016},
\end{itemize}

\noindent
as will be shown in this paper, the following regressors can also be removed
\begin{itemize}
	\item[(iii)] $y^{p}_{k-\tau_y}u^m_{k-\tau_u},~\forall p,\,m$, 
\end{itemize}

\noindent
where $\tau_y$ and $\tau_u$  are any time lags.	$\square$
\end{assump}
%--------------

The steady-state analysis of a model that complies with Assumption~\ref{R2} is done by taking 
$y_k=\bar{y},\,\forall k$, $u_k=\bar{u},\,\forall k$ and, consequently, $\phi_{1,\,k}=u_k-u_{k-1}=0$
$\phi_{2,\,k}={\rm sign}(\phi_{1,\,k})=0,\,\forall k$, thus yielding $\bar{y}=\Sigma_y\bar{y}$, 
where $\Sigma_y$ is the sum of all parameters of all linear output regressors. For $\Sigma_y \neq 1$
the model has a single fixed point at $\bar{y}=0$ with stability domain given by:
\begin{equation}
\label{170919b}
-1 < \qquad \theta_1 \qquad < 1.
\end{equation}

\noindent
If $|\theta_1| < 1$ ($|\theta_1| > 1$), then $\bar{y}=0$ is only non-diverging (diverging) equilibrium and,
as a result, Property~\ref{D1} is not satisfied. In order to solve this problem, we start be reviewing the
following definition.

%------------- Definition 1
\begin{defin}{\rm(Continuum of equilibrium points \cite{Morris2011})}.
	\label{Def1}
	A model has a continuum of equilibrium points if for any constant value of the input
	its corresponding output in steady-state is an equilibrium solution. $\square$
\end{defin}
%-------------

Based on Definition\,\ref{Def1} and the problem aforementioned, the following lemma is stated.

%------------- Lemma 1
\begin{lemma}
	\label{P1}
	Given that Assumption \ref{R2} holds, if $\Sigma_y=1$ is verified, then the identified model
	has a {\it continuum of equilibrium points}\, at steady-state. $\square$
\end{lemma}
%-------------

{\it Proof.} The steady-state analysis of a model that satisfies Assumption~\ref{R2} and Lemma~\ref{P1} yields
$\bar{y}=\bar{y}$ which is trivially true for any value $\bar{y}$. Hence, the model has a
{\it continuum of equilibrium points}\, and Property~\ref{D1} is satisfied.
$\square$

%-----------------------------------------------------------------
\subsection{Quasi-static analysis}
\label{qsa}

The core idea of the framework proposed in \cite{Martins_Aguirre2016} to identify
models with a hysteresis loop is to build a bounding structure ${\cal H}$ made of
sets of equilibria {\it and}\, to ensure that one set is stable during loading and
the other one, during unloading. Such a scenario is effective, but it does not help
to understand models with more complicated structures and with both attracting and
repelling regions in the $u \times y$ plane. This section aims at enlarging the
scenario developed in \cite{Martins_Aguirre2016}.

In quasi-static analysis, it is assumed that the input $u_k$ is a loading-unloading signal
that is much slower than the system dynamics to the point that, at a given time $k$, the 
system will be in a certain {\it attracting}\, region, avoiding any possible {\it repelling}\,
regions. Also, such regions depend on $u_k$, $\phi_{1,\,k}$ and  $\phi_{2,\,k}$.
More specifically, there will be two sets of regions, one for loading and another for unloading.

In quasi-static analysis, we assume that $y_k\approx y_{k-j}=\tilde{y},~j=1,\,2,\ldots,\,n_y$,
such that (\ref{090919a}) is given by
\begin{eqnarray}
\label{090919b}
\tilde{y} & \approx & F^{\ell}\big(\tilde{y}, \,u_{k-1},\cdots,u_{k-n_u}, \,  \phi_{1,\,k-1}, \,\phi_{2,\,k-1} \big), 
\end{eqnarray}

\noindent
which can be usually solved for $\tilde{y}$, especially if higher powers of the output are
not in $F^{\ell}(\cdot)$ \cite{Aguirre_Mendes1996}. This is achieved in practice by removing such
group of terms from the set of candidates. If the model has no inputs, then $\tilde{y}$
coincides with the fixed points. Alternatively, if the inputs are all constant, then 
$\tilde{y}$ is a family of fixed points that depends on the set of constant inputs.

Given the slow input, if $\tilde{y}$ is in an attractive region, then the model output
moves towards $\tilde{y}$. In what follows, $\tilde{y}_{\rm L}^{\rm a}$ and
$\tilde{y}_{\rm U}^{\rm a}$ are, respectively, the solutions to (\ref{090919b}) in
attracting regions under loading and unloading. Likewise, $\tilde{y}_{\rm L}^{\rm r}$
and $\tilde{y}_{\rm U}^{\rm r}$ are their counterparts in repelling regions.
The conditions for $\tilde{y}$ to be attracting is
\begin{eqnarray}
\label{090919c}
\left| {\rm eig} \left( \frac{\partial F^{\ell}(\bm{y},\,u_{k-1},\,\phi_{1,\,k-1},\, \phi_{2,\,k-1})}
{\partial \bm{y}} \right) \right|<1 ,
\end{eqnarray}

\noindent
where $\bm{y}=[y_{k-1}\, \ldots y_{k-n_y}]^T$. This procedure resembles that of determining
the stability of fixed points. Here the Jacobian matrix is not evaluated at fixed points. Hence
we do not speak in terms of stable and unstable fixed points.

\begin{figure}[htb]%[htb]
	\centering{
	\includegraphics[scale=1]{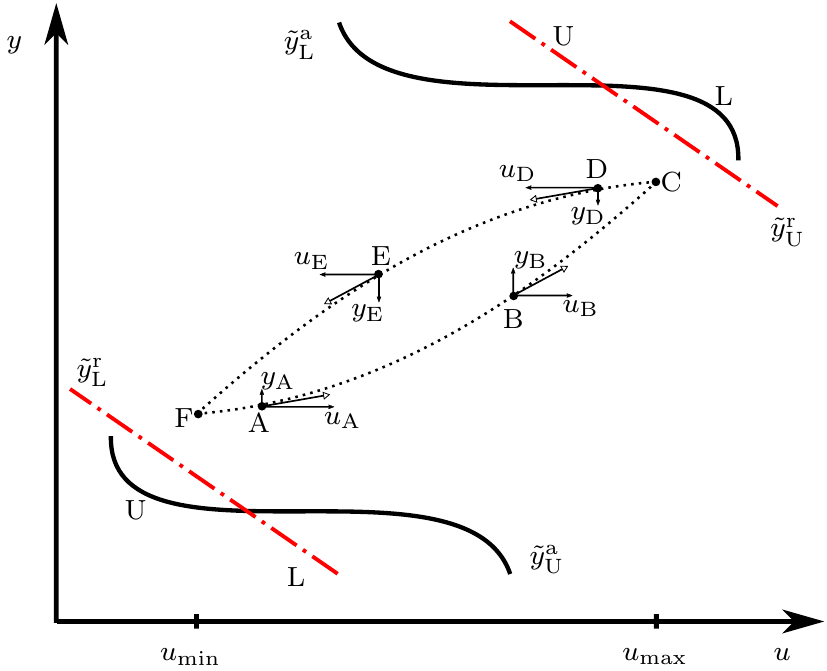} \vspace{-2mm}
		\caption{Schematic representation of hysteresis loop in the $u \times y$ plane.
	Attracting sets are shown in black continuous lines, whereas the repelling
	sets are indicated in red dash-dot. The hysteresis loop is indicated by dotted lines.} % thick lines.
	\label{F2}	
	}
\end{figure}

To illustrate how this helps to understand the formation of a hysteresis loop, consider the schematic
representation in figure\,\ref{F2}. The input is a loading-unloading signal such that
$u_{\rm min} \le u_k \le u_{\rm max},~\forall k$. The sets $\tilde{y}_{\rm L}^{\rm a}$,
$\tilde{y}_{\rm U}^{\rm a}$, $\tilde{y}_{\rm L}^{\rm r}$ and $\tilde{y}_{\rm U}^{\rm r}$
are shown. Consider the point A, which takes place under loading. Given that the system
is under the direct influence of $\tilde{y}_{\rm L}^{\rm r}$, which is responsible for
pushing upwards (see vertical component $y_{\rm A}$), and it is the loading regime, there
is a horizontal component $u_{\rm A}$ (related to the input) that points to the right.
The resulting effect is to pull the system along the loop in the NE direction. The same
can be said for point B; however, at that point the vertical component is the result of
the attracting action of $\tilde{y}_{\rm L}^{\rm a}$.
A similar analysis can be readily done for the unloading regime, given by points D and E.
At the turning points C and F, $\phi_{2,\,k}$ switches from 1 to -1 and from -1 to 1,
respectively. Hence the analysis also switches from using $\tilde{y}_{\rm L}^{\rm a}$
and $\tilde{y}_{\rm L}^{\rm r}$, to using $\tilde{y}_{\rm U}^{\rm a}$ and
$\tilde{y}_{\rm U}^{\rm r}$. This analysis will be useful in
section\,\ref{Section:NumericalResults} to understand the formation of hysteresis loops
in identified models.

As a final remark, it is important to point out that the assumption that the set $\tilde{y}$
comes in two disjoint parts, either for loading or unloading, is a consequence of the solution of
(\ref{090919b}) being rational instead of polynomial. This is useful to analyse models
with more general model structures.

The following example illustrates the application of this analysis and show which constraints
should be considered to comply with Property~\ref{D1}.

%-------------- Example 1
\begin{example}
	\label{Example1}
	Consider the following NARX model that complies with Assumption\,\ref{R2}:
	\begin{eqnarray}
	\label{170919a}
	y_k&{=}\theta_1y_{k-1}+\theta_2\phi_{2,\,k-1}{+}\theta_3\phi_{1,\,k-1}u_{k-1} \nonumber \\
	&{+}\theta_4\phi_{2,\,k-1}\phi_{1,\,k-1}y_{k-1}{+}\theta_5\phi_{1,\,k-1}. 
	\end{eqnarray}
	
	\noindent
	In this case, the constraint $\theta_1=1$ will be used such that, according to Lemma~\ref{P1},
	the resulting model will have a continuum of equilibrium points. This can be achieved
	using estimator (\ref{Eq:ConstrainedParameters_Estimator}) with $c=1$ and $S=[1\,\, 0\,\, 0\,\, 0\,\, 0]$.

	For a more complicated model structure, the constraint in Lemma~\ref{P1} is still
	in the form $1=S\bm{\theta}$ (\ref{Eq:ContrainedParameters_Estimator_Problem}) but with $S$
	having more that one element equal to one, e.g. as shown in \cite{Aguirre2014} to obtain NARX
	models able to reproduce dead-zone and in \cite{Aguirre_etal2004} for a quadratic nonlinearity.

	The quasi-static analysis of model (\ref{170919a}) is performed following the steps provided in section\,\ref{qsa}.
	So rewriting this model as (\ref{090919b}), we have
	\begin{eqnarray}
		\tilde{y} & \approx & \theta_1\tilde{y}{+}\theta_2{\phi}_{2,k-1}{+}\theta_3{\phi}_{1,k-1}u_{k-1}  \nonumber \\
		& ~ &  +\theta_4{\phi}_{2,k-1}{\phi}_{1,k-1}\tilde{y} {+}\theta_5{\phi}_{1,k-1}, \nonumber
	\end{eqnarray}	
	
	\noindent
	which can be described by
	\begin{equation} 
	\label{190919a}
	\tilde{y}({u},{\phi}_{1},{\phi}_{2}) {=}
	\cases{\frac{\theta_2+\theta_3{\phi}_{1}{u}+\theta_5{\phi}_{1}}{1-\theta_1-\theta_4{\phi}_{1}}, {\rm for} \,\,{\phi}_{2}{=}\,\,\,1; \\
	\frac{-\theta_2+\theta_3{\phi}_{1}{u}+\theta_5{\phi}_{1}}{1-\theta_1+\theta_4{\phi}_{1}}, {\rm for} \,\,{\phi}_{2}{=}-1,\\}
	\end{equation}
	
	\noindent
	where the time indices have been omitted for simplicity. Therefore, the solution
	given at the top in (\ref{190919a}) represents the set $\tilde{y}_{\rm L}$, while the bottom
	is the set $\tilde{y}_{\rm U}$.

	To define whether the solutions to (\ref{190919a}) are in the attracting or repelling regions,
	(\ref{090919c}) should be computed for model (\ref{170919a}) as
	\begin{eqnarray}
	\label{190919b}
	-1<&  \theta_1 + \theta_4{\phi}_{2,k-1}{\phi}_{1,k-1} &<1, \nonumber \\
	\frac{-1-\theta_1}{\theta_4\phi_{2,k-1}}<& {\phi}_{1,k-1} &< \frac{1-\theta_1}{\theta_4\phi_{2,k-1}}.
	\end{eqnarray}
	
	\noindent
	Since it is assumed that the known input $u_k$ is a non-zero loading-unloading signal,
	then the conditions (\ref{190919b}) to ensure that the solutions (\ref{190919a}) are in attracting regions
	can be readily verified numerically. In sections \ref{Section:NumericalResults} and
	\ref{Section:ExperimentalResults}, the same analyses will be performed for the identified models. $\square$
\end{example}
%--------------

%==================================================================
\section{Compensator Design} 
\label{Section:Methodology}

The proposed strategies to design compensators based on NARX models are detailed
in this section, starting with some preliminary assumptions. In this paper, a key
point is to investigate if hysteresis in the {\it models}\, estimated according to
section\,\ref{ish} have any impact on the regulation performance.

%-----------------------------------------------------------------
\subsection{Preliminaries} 
\label{Subsection:Compensators:Preliminaries}

\begin{figure}[htb]%[htb]
	\centering{
		\includegraphics[scale=0.93]{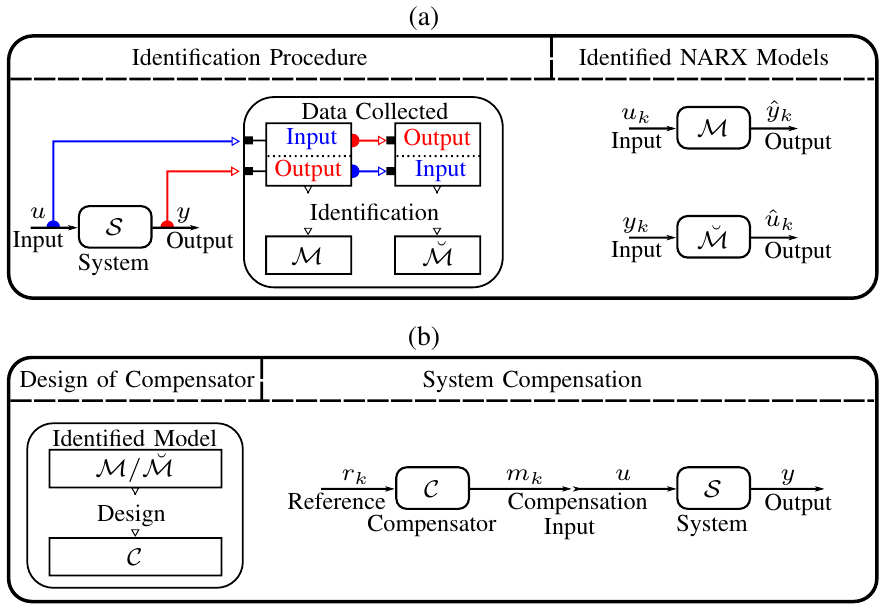} \vspace{-2mm}
		\caption{Compensator design based on identified NARX models. (a)~Model identification, and (b)~compensator design based on identified models.}
		\label{F1}	
	}
\end{figure}

Given a nonlinear system $\mathcal{S}$, the first step is to obtain hysteretic
models for $\mathcal{S}$; see figure\,\ref{F1}(a). To achieve that, two procedures
will be followed. The first one aims at identifying a model $\mathcal{M}$ based on
the direct relationship between $u$ and $y$, whose simulation yields $\hat{y}_k$,
according to section~\ref{Subsection:Compensator:Cms}. The second procedure is based on
the identification of the inverse relationship, in which case a model $\breve{\mathcal{M}}$
is obtained to yield $\hat{u}_k$, as illustrated in figure\,\ref{F1}(a) and following 
section~\ref{Subsection:Compensator:Cci}. In the second step, the identified model is
used to design a compensator $\mathcal{C}$ that yields the compensation signal $m_k$
for a given reference $r_k$; see figure\,\ref{F1}(b).

In this paper,the following additional assumptions are made for NARX models\,(\ref{090919a}).

%-------------- Remark 3 -> (2 in the text)
\begin{remark}
	\label{R3}
	For compensation design, $y_k$ can be replaced by $r_k$, and $u_k$ by $m_k$,
	respectively, in the models $\mathcal{M}$ and $\breve{\mathcal{M}}$. $\square$
\end{remark}
%-------------- 

The motivation behind this is that $y_k$ should ideally be equal to $r_k$ under
compensation, that is, when $m_k$ is used as an input to the dynamical system.

In what follows, the main idea is to use an identified model to determine
the compensation input $m_{k-\tau_{\rm d}+1}$.

%-----------------------------------------------------------------
\subsection{Model-Based Compensation} 
\label{Subsection:Compensator:Cms}

The aim here is to specify a general model structure for $\mathcal{M}$ in order to
find $m_k$ analytically from this model. To achieve that, the following assumptions
are needed.

%-------------- Assumption 1 -> (2 in the text)
\begin{assump} \label{A1}
	Assume that:
	(i)~the only regressor involving $u_{k-\tau_{\rm d}}$ is linear;
	(ii)~$n_u>\tau_{\rm d}$;
	(iii)~the compensation signal $m_k$ is known up to time $k-\tau_{\rm d}$;
	and (iv)~the reference signal $r_k$ is known up to time $k+1$. $\square$
\end{assump}
%-------------- 

Assumption~\ref{A1} imposes conditions on the selection of the model structure. 
Note that (i)~ensures that $u_{k-\tau_{\rm d}}$ can be isolated in the identified models;
(ii)~allows that input terms with a delay longer than $\tau_{\rm d}$ to be regressors
in the identification procedure; and the other constraints guarantee that the control
action can be computed from known values. Therefore, the model $\mathcal{M}$ is rewritten as
\begin{equation}
\label{Eq:NARX_Extended_Compensator:Cms:Initial}
A(q)y_k {=} B(q)u_{k}{+}f\big(y_{k-1},{\cdots},y_{k-n_{y}},u_{k-\tau_{\rm d}-1},{\cdots}, 
u_{k-n_u}\big)\!,
\end{equation}

\noindent
where $q^{-1}$ is the backward time-shift operator such that $q^{-1}u_k{=}u_{k-1}$, 
and the linear regressors are grouped in $A(q)y_k$ and $B(q)u_{k}$ with
\begin{eqnarray}
A(q) &=& 1 - a_1q^{-1} -  a_2q^{-2} - \cdots - a_{n_{y}}q^{-n_{y}}, \label{Eq:NARX_Extended_Compensator:Cms:A_q}\\
B(q) &=& b_{\tau_{\rm d}}q^{-\tau_{\rm d}} + \underbrace{b_{\tau_{\rm d}+1}q^{-\tau_{\rm d}-1} + \cdots + b_{n_{u}}q^{-n_{u}}}_{B^*(q)}, \label{Eq:NARX_Extended_Compensator:Cms:Bu_q}
%&=& b_{\tau_{\rm d}}q^{-\tau_{\rm d}} + B^*(q), \label{Eq:NARX_Extended_Compensator:Cms:Bu_q}
\end{eqnarray}

\noindent
and $f(\cdot)$ includes all the nonlinear terms and possibly the constant term of the
NARX model (\ref{090919a}). Using (\ref{Eq:NARX_Extended_Compensator:Cms:Bu_q}),
model (\ref{Eq:NARX_Extended_Compensator:Cms:Initial}) can be rewritten as
\begin{eqnarray}
\label{Eq:NARX_Extended_Compensator:Cms:Intermediary}
A(q)y_k & = & b_{\tau_{\rm d}} u_{k-\tau_{\rm d}}{+}B^*(q)u_{k}{+}f\big(y_{k-1}, \cdots,
y_{k-n_{y}},\nonumber \\ 
& ~& \hspace{2.0cm} u_{k-\tau_{\rm d}-1}, \cdots, u_{k-n_u} \big).
\end{eqnarray}

\noindent
From Remark~\ref{R3}, we have 
\begin{eqnarray}
\label{Eq:NARX_Extended_Compensator:Cms_Used_Remark}
A(q)r_{k{+}1} & = & b_{\tau_{\rm d}} m_{k-\tau_{\rm d}+1} {+}B^*(q)m_{k{+}1}{+}f\big(r_{k},\cdots,\nonumber \\ 
& ~& \hspace{0.5cm} r_{k-n_{y}{+}1}, m_{k-\tau_{\rm d}},\cdots,m_{k-n_u{+}1}\big), 
\end{eqnarray}

\noindent
which, for convenience, has been written an instant of time ahead, i.e. $k \rightarrow k+1$.

From the Assumption\,\ref{A1}, the compensation input can be obtained from
(\ref{Eq:NARX_Extended_Compensator:Cms_Used_Remark}) as
\begin{eqnarray}
\label{Eq:Cms:InputCompensation}
m_{k-\tau_{\rm d}+1}  & = & \frac{1}{b_{\tau_{\rm d}} } \Big[A(q)r_{k{+}1}-B^*(q)m_{k{+}1}-f\big(r_k,\cdots,\nonumber\\
& ~& \hspace{0.5cm} r_{k-n_y{+}1}, m_{k-\tau_{\rm d}},\cdots,m_{k-n_u{+}1}\big)\Big].
\end{eqnarray}

To illustrate the application of this strategy, assume that the constraints discussed
in section\,\ref{ish} and Assumption\,\ref{A1} are verified in the identification procedure.

%-------------- Example 2
\begin{example}
	\label{Example2}
	Consider the NARX model described by
	\begin{eqnarray}
	\label{190919c}
	y_k&{=}\theta_1y_{k-1}{+}\theta_2{\phi}_{2,\,k-2}{+}\theta_3{\phi}_{1,\,k-2}u_{k-2} \nonumber \\
	&{+}\theta_4{\phi}_{2,\,k-2}{\phi}_{1,\,k-2}y_{k-1}{+}\theta_5{\phi}_{1,\,k-1}.
	\end{eqnarray}
	
	\noindent
	Since $\phi_{1,\,k}{=}u_k-u_{k-1}$ and $\phi_{2,\,k} {=} {\rm sign}(\phi_{1,\,k})$, we have
	\begin{eqnarray}
	%\label{190919d}
	y_k&{=}\theta_1y_{k-1}{+}\theta_2{\rm sign}(u_{k-2}-u_{k-3}){+}\theta_3[u_{k-2}-u_{k-3}]u_{k-2} \nonumber \\
	&{+}\theta_4{\rm sign}(u_{k-2}{-}u_{k-3})[u_{k-2}{-}u_{k-3}]y_{k-1}{+}\theta_5[u_{k-1}{-}u_{k-2}], \nonumber
	\end{eqnarray}
	
	\noindent
	which is in the form (\ref{Eq:NARX_Extended_Compensator:Cms:Initial}) and, therefore,
	\begin{equation}
	\label{200919a}
	A(q)y_k {=} B(q)u_{k}{+}f\big(y_{k-1},u_{k-2},u_{k-3},{\rm sign}(u_{k-2}-u_{k-3})\big),
	\end{equation}
	
	\noindent
	where
	\begin{eqnarray}
	A(q) &=& 1 - \theta_1q^{-1}, \label{200919b}\\
	B(q) &=& \theta_5q^{-1} - \theta_5q^{-2}, \label{200919c} \\
	f(\cdot) &=& \theta_2{\rm sign}(u_{k-2}-u_{k-3}){+}\theta_3[u_{k-2}-u_{k-3}]u_{k-2} \nonumber \\
	&~&{+}\theta_4{\rm sign}(u_{k-2}-u_{k-3})[u_{k-2}-u_{k-3}]y_{k-1}.\label{200919d}
	\end{eqnarray}
	
	From Remark\,\ref{R3}, the model (\ref{200919a}) is recast as
	\begin{eqnarray}
	\label{200919f}
	A(q)r_{k{+}1} {=} \theta_5m_{k}{-}&\theta_5m_{k-1}{+}f\big(r_{k},m_{k-1},m_{k-2},\nonumber \\
	&{\rm sign}(m_{k-1}-m_{k-2})\big),
	\end{eqnarray}
	
	\noindent
	and
	\begin{eqnarray}
	\label{200919g}
	m_k & = &\frac{1}{\theta_5} \Big[A(q)r_{k{+}1}{+}\theta_5m_{k-1} \nonumber \\
	& ~& \hspace{0.2cm} -f\big(r_k,m_{k-1},m_{k-2},{\rm sign}(m_{k-1}{-}m_{k-2})\big)\Big], \nonumber \\
	& = &\frac{1}{\theta_5}\Big[r_{k{+}1}-\theta_1r_{k}{+}\theta_5m_{k-1} \nonumber \\
	& ~& \hspace{0.2cm}-\theta_2{\rm sign}(m_{k-1}{-}m_{k-2})-\theta_3[m_{k-1}{-}m_{k-2}]m_{k-1} \nonumber \\
	& ~& \hspace{0.2cm}-\theta_4{\rm sign}(m_{k-1}{-}m_{k-2})[m_{k-1}{-}m_{k-2}]r_{k} \Big],
	\end{eqnarray}
	
	\noindent
	which is computed due to Assumption~\ref{A1}. $\square$
\end{example}
%--------------

%-----------------------------------------------------------------
\subsection{Compensation Based on Compensator Identification} 
\label{Subsection:Compensator:Cci}

Here, the strategy is to identify NARX models $\breve{\mathcal{M}}$ that
are able to describe the inverse relationship between the input $u$ and
output $y$ signals of the system $\mathcal{S}$. The advantage of this
strategy is that the compensator $\mathcal{C}$ is obtained directly
from $\breve{\mathcal{M}}$, according to the variable solutions
presented in Remark\,\ref{R3}. However, some issues related to the
identification procedure of these models need to be addressed.
For nomenclature simplicity, in this section, we assume that $\tau_{\rm d}=1$.

For the inverse model $\breve{\mathcal{M}}$, the output $\hat{u}_k$ depends on $y_k$.
Hence in order to avoid the lack of causality, $y_k$ should be delayed by $\tau_{\rm s}$ 
time steps with respect to $u_k$, yielding \cite{Xia2003}:
\begin{equation}
\hat{u}_k {=} \breve{F}\big(\hat{u}_{k-1},\cdots,\hat{u}_{k-n_{u}},y_{k-1{+}\tau_{\rm s}},\cdots,y_{k-n_y+\tau_{\rm s}}\big), \label{Eq:NARX_Extended_Model:Cci}
\end{equation}

\noindent
where $\breve{F}(\cdot)$ is the inverse nonlinear function and $\hat{u}_k \in \mathbb{R}$
and $y_k \in \mathbb{R}$ are related as shown in figure\,\ref{F1}(a). It should be noted
that $\tau_{\rm s} \ge \tau_{\rm d}+1$, where usually the equality is preferred. Similar
ways to avoid noncausal models can be found in the literature
\cite{Rakotondrabe2011,L_Junior_etal2019}.

%-------------- Assumption 2 -> (3 in the text)
\begin{assump} 
	\label{A2}
	Assume that:
	(i)~there is at least one regressor of the output $(y_k)^j$ for $j \geq1$;
	(ii)~the compensation signal $m_k$ is known up to time $k-1$;
	and (iii)~the reference signal $r_k$ is known up to time $k-1+\tau_{\rm s}$. $\square$
\end{assump}
%-------------- 

Assumption~\ref{A2} should be imposed during the structure selection of the inverse model
$\breve{\mathcal{M}}$. Note that (i)~ensures that there is at least
one input signal $y_k$ in the identified models; (ii) and (iii) ensure that the compensation
input $m_k$ to be computed at time $k$ is the only unknown variable. Given Assumption~\ref{A2}
and Remark~\ref{R3}, the compensation signal $m_k$ can be obtained directly from $\breve{\mathcal{M}}$ as
\begin{equation}
\label{Eq:Cci:InputCompensation}
m_k {=} \breve{F}\big(m_{k-1},\cdots,m_{k-n_{u}},r_{k-1{+}\tau_{\rm s}},\cdots,r_{k-n_y+\tau_{\rm s}} \big). 
\end{equation}

%As a final remark, attention must be drawn to the importance of parameter $\tau_{\rm s}$ that should
%be applied to the identification and validation data sets. Hence, this parameter can be omitted
%in the inverse model (\ref{Eq:NARX_Extended_Model:Cci}) for the case where the data collected
%from the system are rearranging a priori, but it must be taken into account in the compensation
%model (\ref{Eq:Cci:InputCompensation}), as will be seen in sections \ref{Section:NumericalResults}
%and \ref{Section:ExperimentalResults}.

%==================================================================
\section{Numerical Results} 
\label{Section:NumericalResults}

This section identifies models to predict the behavior of a hysteretic system from simulated data,
and evaluates the performance of these models in predicting dynamics and compensating the nonlinearity
of the simulated system.

%-----------------------------------------------------------------
\subsection{Identification of a bench test system}

Consider the piezoelectric actuator with hysteretic nonlinearity modeled
by the Bouc-Wen model \cite{Wen1976} and whose mathematical model is given
by \cite{Rakotondrabe2011}
\begin{equation}
\label{Eq:NumericalResults:System}
\cases{\dot{h}(t) = ~A\dot{u}(t) - \beta|\dot{u}(t)|h(t) - \gamma\dot{u}(t)|h(t)|,\\
y(t) =~d_{\rm p} u(t) - h(t),\\}
\end{equation}

\noindent
where $y(t)$ is the displacement, $u(t)$ is the voltage applied to the actuator, 
$d_{\rm p}\hspace{-0.05cm}=\hspace{-0.05cm}1.6\, {\rm \frac{\mu m}{V}}$ is the
piezoelectric coefficient, $h(t)$ is the hysteretic nonlinear term and
$A\hspace{-0.05cm}=\hspace{-0.05cm}0.9\, {\rm \frac{\mu m}{V}}$, 
$\beta\hspace{-0.05cm}=\hspace{-0.05cm}0.008\, {\rm V^{-1}}$ and
$\gamma\hspace{-0.05cm}=\hspace{-0.05cm}0.008\, {\rm V^{-1}}$ are parameters
that determine the shape and scale of the hysteresis loop.

\begin{figure}[htb]%[htb]
	\centering{
		\includegraphics[scale=0.97]{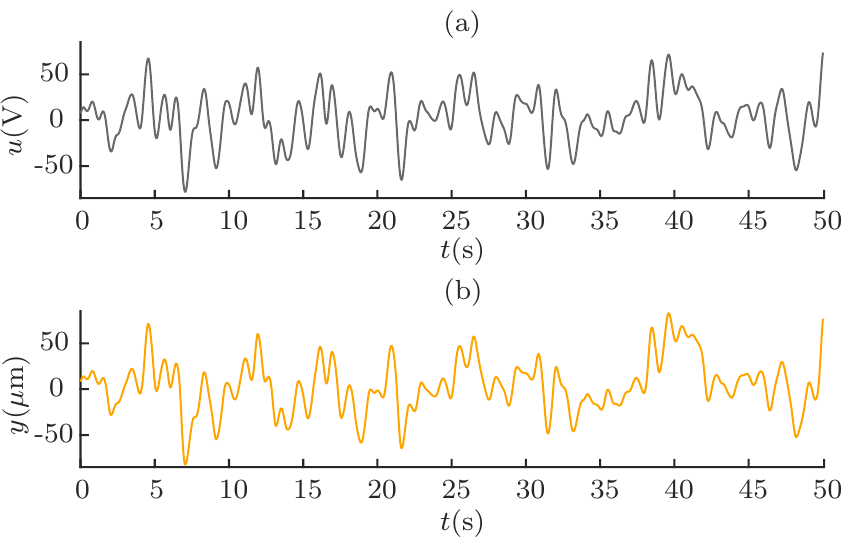} \vspace{-2mm}
		\caption{Signals used to identify system (\ref{Eq:NumericalResults:System}). (a)~excitation, and (b)~simulated output.} % thick lines.
		\label{Fig:NumericalResults:Input_Output_Indet}	
	}
\end{figure}

Model (\ref{Eq:NumericalResults:System}) was integrated numerically using a fourth-order
Runge-Kutta method with integration step $\delta t=0.001\,{\rm s}$.
The excitation signal was generated by low-pass
filtering a white Gaussian noise \cite{Martins_Aguirre2016}. In this work, a fifth-order
low-pass Butterworth filter with a cutoff frequency of $1$\,Hz was used;
see figure\,\ref{Fig:NumericalResults:Input_Output_Indet}(a). The sampling time is set to
$T_{\rm s}=\delta t=0.001\,{\rm s}$ and a frequency of $1$\,Hz is chosen
to validate the identified models \cite{Rakotondrabe2011}. The data sets are
$50\,{\rm s}$ long $(N=50000)$.

The constraints defined in section\,\ref{ish} are here considered to build NARX models
for system (\ref{Eq:NumericalResults:System}). Input and output signals are shown in
figure\,\ref{Fig:NumericalResults:Input_Output_Indet}. In addition to the monomial
regressors in $u_k$ and $y_k$, the following regressors are also used: $\phi_{1,\,k}=u_k-u_{k-1}$
and the sign of this first difference $\phi_{2,\,k} = {\rm sign}(\phi_{1,\,k})$.
The maximum nonlinear degree for regressors is cubic, $\ell=3$, and the maximum delays
are $n_y=n_u=1$. This choice is based on the fact that discrete models of hysteresis
that have only unit delayed regressors typically perform well and result in models with
simple structures, which are advantageous for model-based control \cite{Martins_Aguirre2016,L_Junior_etal2019}.

The model structure is selected using the ERR criterion to rank the regressors according to importance
and the AIC determine the final number of model terms. In this case, the standard least squares solution
is used for parameter estimation. Also, as proposed in section\,\ref{ish}, to obtain models that describe
some features of hysteresis, the constrained parameter estimation was used in order to comply with the
condition established in Property~\ref{D1}.

%-----------------------------------------------------------------
\subsubsection{Estimating $\mathcal{M}$.}

In this example we take the following metaparameters: $n_u=2$ which is the
smallest value that complies with Assumption\,\ref{A1}-(ii); while $n_y$ and $\ell$
maintain the values which were determined above. Using the data shown in
figure~\ref{Fig:NumericalResults:Input_Output_Indet} and considering
Assumption\,\ref{R2}, the following model structure is obtained
\begin{eqnarray}
\label{Eq:NumericalResults:IdentifiedModel:Cms}
y_k&=\theta_1y_{k-1}+\theta_2\phi_{1,\,k-1} +\theta_3\phi_{2,\,k-2}\phi_{1,\,k-2}u_{k-2} \nonumber\\
&+\theta_4\phi_{2,\,k-2}\phi_{1,\,k-2}y_{k-1}+\theta_5\phi_{1,\,k-2}u^2_{k-2}  \nonumber\\
&+\theta_6\phi_{1,\,k-2}u_{k-2}y_{k-1}, 
\end{eqnarray}

\noindent
where $\theta_i$ for $i=1,\cdots,6$ are estimated by least squares.

In steady-state, we have $y_k=\bar{y},\,\forall k$, $\phi_{1,\,k}=0,\,\forall k$;
hence, the resulting expression is $\bar{y}=\theta_1\bar{y}$. Therefore, based on
Lemma\,\ref{P1} and Example\,\ref{Example1}, for model
(\ref{Eq:NumericalResults:IdentifiedModel:Cms}) to fulfill Property\,\ref{D1},
the constraint $\Sigma_y=\theta_1=1$ should be imposed. This can be done using
(\ref{Eq:ConstrainedParameters_Estimator}) with the constraint written as:
\begin{equation}
\label{Eq:NumericalResults:Constraints}
c = 1;\qquad
S {=} [1 ~~ 0 ~~ 0 ~~ 0 ~~ 0 ~~ 0]. 
\end{equation}

\noindent
Hence, the parameter values estimated by the constrained least squares estimator
(\ref{Eq:ConstrainedParameters_Estimator}) are shown in Table\,\ref{Tab1}.

\begin{table}[htb]%[htb]
	\centering
	\caption{Model parameters obtained with
		(\ref{Eq:ConstrainedParameters_Estimator}) and (\ref{Eq:NumericalResults:Constraints}).} %\vspace{0.1cm}
	\label{Tab1} %\setlength\tabcolsep{4pt} % default value: 6pt
	\begin{tabular}{c|c c}
		%\hline
		\textbf{Model}   &   \multicolumn{2}{c}{Values}  \\ \hline
		\multirow{3}{*}{(\ref{Eq:NumericalResults:IdentifiedModel:Cms})} & $\theta_1{=}1.00$ & $\theta_2{=}0.77$\\
		&  $\theta_3{=}1.44\times10^{-2}$ &  $\theta_4{=}-9.60\times10^{-3}$ \\
		& $\theta_5{=}3.15\times10^{-4}$ & $\theta_6{=}-2.47\times10^{-4}$  \\ \hline
		\multirow{3}{*}{(\ref{Eq:NumericalResults:IdentifiedModel:Cci})} & $\theta_1{=}1.00$ & $\theta_2{=}1.27$\\
		&  $\theta_3{=}-2.13\times10^{-2}$ &  $\theta_4{=}1.37\times10^{-2}$ \\
		& $\theta_5{=}-1.07\times10^{-5}$ & $\theta_6{=}7.99\times10^{-6}$  \\
	\end{tabular}
\end{table}

Next, a quasi-static analysis of the identified model is performed as
discussed in section\,\ref{qsa} and illustrated in Example\,\ref{Example1}.
First, we write for (\ref{Eq:NumericalResults:IdentifiedModel:Cms}) the
corresponding to (\ref{090919b}) as
\begin{eqnarray}
\tilde{y} & \approx & \theta_1\tilde{y}+\theta_2{\phi}_{1,k-1} +\theta_3{\phi}_{2,k-2}{\phi}_{1,k-2}u_{k-2} \nonumber \\
& ~ & {+}\theta_4{\phi}_{2,k-2}{\phi}_{1,k-2}\tilde{y}\!+\!\theta_5{\phi}_{1,k-2}u_{k-2}^2\!+\!\theta_6{\phi}_{1,k-2}u_{k-2}\tilde{y}, \nonumber 
\end{eqnarray}

\noindent
yielding
\begin{equation} 
\label{Eq:NumericalResults:QuasiStatic:EP:Cms}
\tilde{y}({u},{\phi}_{1},{\phi}_{2}) {=} \cases{
\frac{\theta_2{\phi}_{1}+\theta_3{\phi}_{1}{u}+\theta_5{\phi}_{1}{u}^2}{1-\theta_1-\theta_4{\phi}_{1}-\theta_6{\phi}_{1}{u}}, {\rm for} \,\,{\phi}_{2}{=}\,\,\,1; \\
\frac{\theta_2{\phi}_{1}-\theta_3{\phi}_{1}{u}+\theta_5{\phi}_{1}{u}^2}{1-\theta_1+\theta_4{\phi}_{1}-\theta_6{\phi}_{1}{u}}, {\rm for} \,\,{\phi}_{2}{=}-1,} 
\end{equation}

\noindent
where the time indices have been omitted for brevity.

The top expression in (\ref{Eq:NumericalResults:QuasiStatic:EP:Cms}) gives the
set $\tilde{y}_{\rm L}$, while the bottom one, $\tilde{y}_{\rm U}$. Computing the
derivative of (\ref{Eq:NumericalResults:IdentifiedModel:Cms}) with respect to
$y_{k-1}$ and using (\ref{090919c}), we obtain
\begin{small}
	\begin{eqnarray}
	\label{d090919}
	-1< \quad \theta_1 + \theta_4\phi_{1,k-2}{\phi}_{2,k-2}+\theta_6{\phi}_{1,k-2}u_{k-2} \quad <1, \nonumber \\
	\frac{-1-\theta_1-\theta_4\phi_{1,k-2}{\phi}_{2,k-2}}{\theta_6\phi_{1,k-2}}< u_{k-2} < \frac{1-\theta_1-\theta_4\phi_{1,k-2}{\phi}_{2,k-2}}{\theta_6\phi_{1,k-2}}.
	\end{eqnarray}
\end{small}

\begin{figure}[htb]%[htb]
	\centering{
		\includegraphics[scale=0.97]{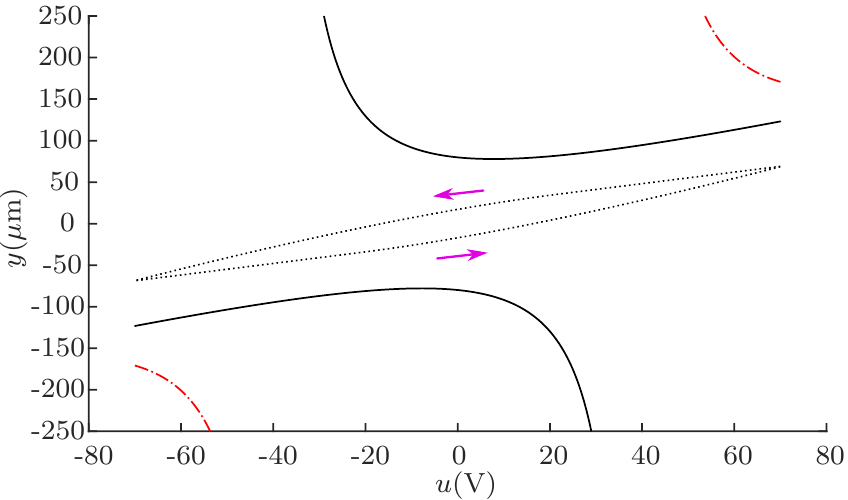} \vspace{-2mm}
		\caption{Results of quasi-static analysis for model~(\ref{Eq:NumericalResults:IdentifiedModel:Cms}) with input $u_k{=}70\sin(2\pi k)\,{\rm V}$. The hysteresis loop indicated with (\textcolor{black}{$\cdots$})~is a result of the interaction of 
			(\textcolor{black}{---})~attracting ($\tilde{y}_{\rm L}^{\rm a}$, $\tilde{y}_{\rm U}^{\rm a}$) and
			(\textcolor{red}{-$\,\cdot\,$-})~repelling ($\tilde{y}_{\rm L}^{\rm r}$, $\tilde{y}_{\rm U}^{\rm r}$) sets.
			(\textcolor{magenta}{\chemarrow}) indicates the orientation of the hysteresis loop.} % thick lines.
		\label{Fig4}	
	}
\end{figure}

Taking ${\phi}_{2,k-2}{=}1$ or ${\phi}_{2,k-2}{=}-1$, the conditions for attracting
regions under load or unloading, respectively, are obtained. Considering the parameter
values presented in Table\,\ref{Tab1} and a loading-unloading input signal, the points
(\ref{Eq:NumericalResults:QuasiStatic:EP:Cms}) and their attraction conditions
(\ref{d090919}) are computed numerically and shown in figure\,\ref{Fig4}.
Figure\,\ref{Fig4} should be compared to figure~\ref{F2}, whose main elements are analogous.
Hence, in this way it is possible to see how model (\ref{Eq:NumericalResults:IdentifiedModel:Cms})
is able to describe the hysteresis nonlinearity.

\begin{figure}[htb]%[htb]
	\centering{
		\includegraphics[scale=0.98]{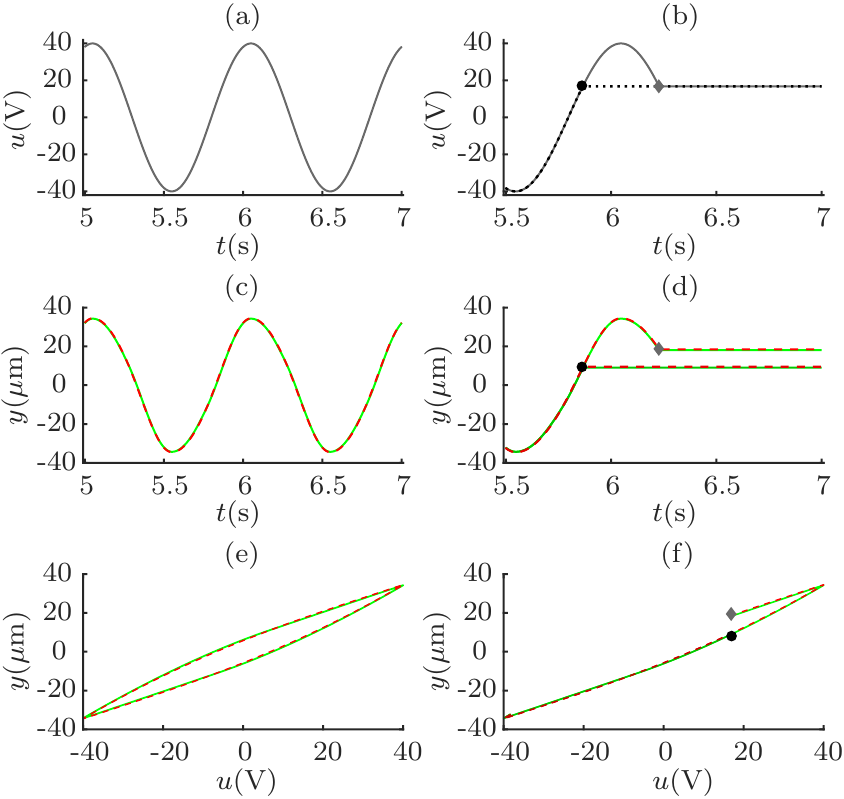} \vspace{-2mm}
		\caption{Free-run simulation of model (\ref{Eq:NumericalResults:IdentifiedModel:Cms}). This figure is arranged in columns, which have: (a) sinusoidal input of voltage $u_k{=}40\sin(2\pi k)\,{\rm V}$ and in (b) the case where this input becomes constant during a loading ($\bullet$) and unloading (\textcolor{gray}{$\blacklozenge$}) regime with the final value of $16.8\,{\rm V}$, its temporal responses are shown in (c) and (d) while the hysteresis loops are in (e) and (f), respectively. (\textcolor{green}{---}) represents the measured data and (\textcolor{red}{- -}) is the estimated output of the model. The full records have $N=50000$ data points.} % thick lines.
		\label{Fig:NumericalResults:Valid_Model_C_ms}	
	}
\end{figure}

Model (\ref{Eq:NumericalResults:IdentifiedModel:Cms}) is simulated with a loading-unloading input
(see left side of figure\,\ref{Fig:NumericalResults:Valid_Model_C_ms}) and, in cases where the input
becomes constant, either during loading or unloading (see right side of figure\,\ref{Fig:NumericalResults:Valid_Model_C_ms}),
{\it the system remains at the corresponding point of the hysteresis loop.}
This is a direct consequence of using Lemma\,\ref{P1}. This feature is not generally present in
identified models found in the literature.

The mean absolute percentage error ({\rm MAPE})
\begin{equation}
{\rm MAPE} = \frac{100 \sum_{k=1}^{N} |y_k-\hat{y}_k|}{N|\max(\bm{y}) - \min(\bm{y})|}, \label{Eq:MAPE}
\end{equation}

\noindent
computed for the case in figure~\ref{Fig:NumericalResults:Valid_Model_C_ms}(c),
is shown in Table\,\ref{Tabel:NumericalResults:IdentifiedModels:PerformanceIndex}.

%-----------------------------------------------------------------
\subsubsection{Estimating $\breve{\mathcal{M}}$.}

The identified model that complies with Assumptions\,\ref{R2} and \ref{A2} is given by
\begin{eqnarray}
\label{Eq:NumericalResults:IdentifiedModel:Cci}
\hat{u}_k&=\theta_1\hat{u}_{k-1}+\theta_2\breve{\phi}_{1,\,k-1} +\theta_3\breve{\phi}_{2,\,k-1}\breve{\phi}_{1,\,k-1}\hat{u}_{k-1} \nonumber\\
&+\theta_4\breve{\phi}_{2,\,k-1}\breve{\phi}_{1,\,k-1}y_{k-1}+\theta_5\breve{\phi}_{2,\,k-1}y_{k-1}\hat{u}_{k-1}  \nonumber\\
&+\theta_6\breve{\phi}_{2,\,k-1}y^2_{k-1}, 
\end{eqnarray}

\noindent
where $\breve{\phi}_{1,\,k}=y_k-y_{k-1}$, $\breve{\phi}_{2,\,k} = {\rm sign}(\breve{\phi}_{1,\,k})$,
$\hat{u}_k$ is the estimated input (model output), and $y_k$ is the output of system (\ref{Eq:NumericalResults:System})
(model input).

Note that the regressors of (\ref{Eq:NumericalResults:IdentifiedModel:Cms})
and of (\ref{Eq:NumericalResults:IdentifiedModel:Cci}) are different. In both cases, the regressors
are automatically chosen from the pool of candidates using the ERR criterion. Nevertheless, also
for (\ref{Eq:NumericalResults:IdentifiedModel:Cci}), the steady-state analysis yields
$\bar{\hat{u}}{=}\theta_1\bar{\hat{u}}$, which is similar to the result found for model
(\ref{Eq:NumericalResults:IdentifiedModel:Cms}). Proceeding as before, the constrained least squares
estimated parameters are shown in Table\,\ref{Tab1}.

\begin{figure}[htb]%[htb]
	\centering{
		\includegraphics[scale=0.97]{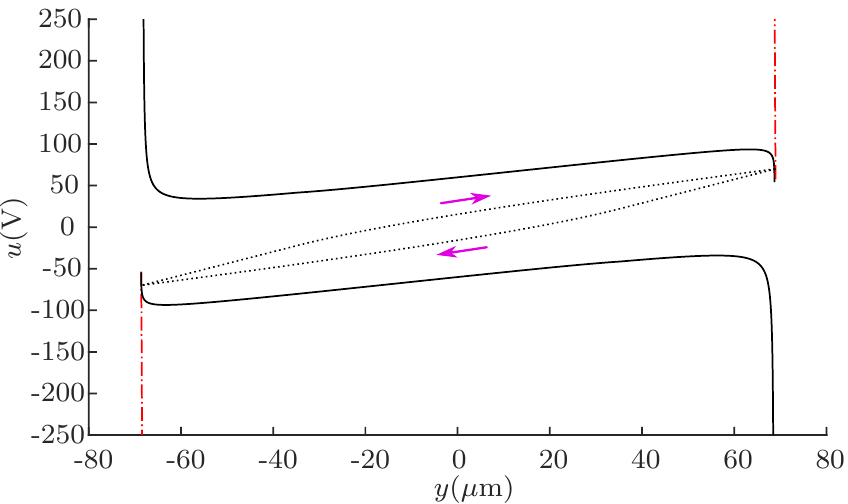} \vspace{-2mm}
		\caption{Results of quasi-static analysis for model (\ref{Eq:NumericalResults:IdentifiedModel:Cci}). For meaning of line
			patterns refer to captions of figure~\ref{F2} and of figure~\ref{Fig4}.} % thick lines.
		\label{Fig:NumericalResults:Analysis_Quasi_Static_C_ci}
	}
\end{figure}

Consider now the quasi-static analysis of model (\ref{Eq:NumericalResults:IdentifiedModel:Cci}).
The formation of the hysteresis loop for this model is shown in figure~\ref{Fig:NumericalResults:Analysis_Quasi_Static_C_ci}. 
Interestingly, the ability of model (\ref{Eq:NumericalResults:IdentifiedModel:Cms}) to describe hysteresis is
also present in model (\ref{Eq:NumericalResults:IdentifiedModel:Cci}). The main difference between them is the
orientation of the hysteresis loop, as discussed in \cite{Gu_etal2012} and illustrated in
figures~\ref{Fig4} and \ref{Fig:NumericalResults:Analysis_Quasi_Static_C_ci}.

Table\,\ref{Tabel:NumericalResults:IdentifiedModels:PerformanceIndex} shows the
prediction performance of models (\ref{Eq:NumericalResults:IdentifiedModel:Cms})
and (\ref{Eq:NumericalResults:IdentifiedModel:Cci}) are similar. In addition, 
Table\,\ref{Tabel:NumericalResults:IdentifiedModels:PerformanceIndex} reports
the prediction performance of a black-box NARX polynomial model. The results
obtained for (\ref{Eq:NumericalResults:IdentifiedModel:Cci}) are similar to
those shown in figure\,\ref{Fig:NumericalResults:Valid_Model_C_ms}
and are omitted for brevity.

\begin{table}[htb]%[htb]
	\centering
	\caption{Performance of the modeling step. Simulation results.} %Performance index of the identified models for a bench test.
	\label{Tabel:NumericalResults:IdentifiedModels:PerformanceIndex} %\setlength\tabcolsep{6.0pt} % default value: 6pt
	\begin{tabular}{c|c|c}
		%\hline
		\textbf{Design Strategy }&\textbf{Model}&{\rm MAPE} \\ \hline 
		Section\,\ref{Subsection:Compensator:Cms} & (\ref{Eq:NumericalResults:IdentifiedModel:Cms}) & $0.398$ \\ \hline
		Section\,\ref{Subsection:Compensator:Cci} & (\ref{Eq:NumericalResults:IdentifiedModel:Cci}) & $0.413$ \\ \hline
		Black-box							   &  not shown										 & $1.494$ \\
	\end{tabular}
\end{table}

%-----------------------------------------------------------------
\subsection{Compensation of a bench test system}

Next, the models identified in the previous section are used to
design compensators using the procedure illustrated in figure\,\ref{F1}(b).

%that are employed to mitigate the nonlinearity inherent
%in the system~(\ref{Eq:NumericalResults:System}) (see figure\,\ref{F1}(b)).

%\pagebreak

%-----------------------------------------------------------------
\subsubsection{Design of the compensation input signals.}

Applying the steps described in section\,\ref{Subsection:Compensator:Cms} to model
(\ref{Eq:NumericalResults:IdentifiedModel:Cms}), the following compensation signal
is obtained
\begin{eqnarray}
\label{Eq:NumericalResults:InputCompensation:Cms}
m_k{=}\frac{1}{\theta_2}&\Big[r_{k{+}1}-\theta_1r_{k}{+}\theta_2m_{k-1} \nonumber\\
&{-}[\theta_3m_{k-1}{+}\theta_4r_{k}]{\rm sign}(m_{k-1}{-}m_{k-2})[m_{k-1}{-}m_{k-2}]  \nonumber\\
&{-}[\theta_5m^2_{k-1}{+}\theta_6m_{k-1}r_{k}][m_{k-1}-m_{k-2}]\Big]. 
\end{eqnarray}

Similarly, using the method described in section\,\ref{Subsection:Compensator:Cci} to
identify the inverse model (\ref{Eq:NumericalResults:IdentifiedModel:Cci}), the following
compensation signal is obtained
\begin{eqnarray}
\label{Eq:NumericalResults:InputCompensation:Cci}
\breve{m}_k&=\theta_1\breve{m}_{k-1}{+}\theta_2[r_{k{+}1}-r_{k}] \nonumber\\
&+[\theta_3\breve{m}_{k-1}+\theta_4r_{k{+}1}]{\rm sign}(r_{k{+}1}-r_{k})[r_{k{+}1}-r_{k}]\nonumber\\
&+[\theta_5r_{k{+}1}\breve{m}_{k-1}+\theta_6r^2_{k{+}1}]{\rm sign}(r_{k{+}1}-r_{k}). 
\end{eqnarray}

Since the parameters of compensators (\ref{Eq:NumericalResults:InputCompensation:Cms}) and
(\ref{Eq:NumericalResults:InputCompensation:Cci}) have been estimated (Table\,\ref{Tab1}),
and Assumptions\,\ref{A1} and \ref{A2} are satisfied, the compensation inputs $m_k$ and
$\breve{m}_k$ can be computed.

%-----------------------------------------------------------------
\subsubsection{Evaluating the performance of the compensation.}

\begin{figure}[htb]%[htb]
	\centering{
		\includegraphics[scale=0.97]{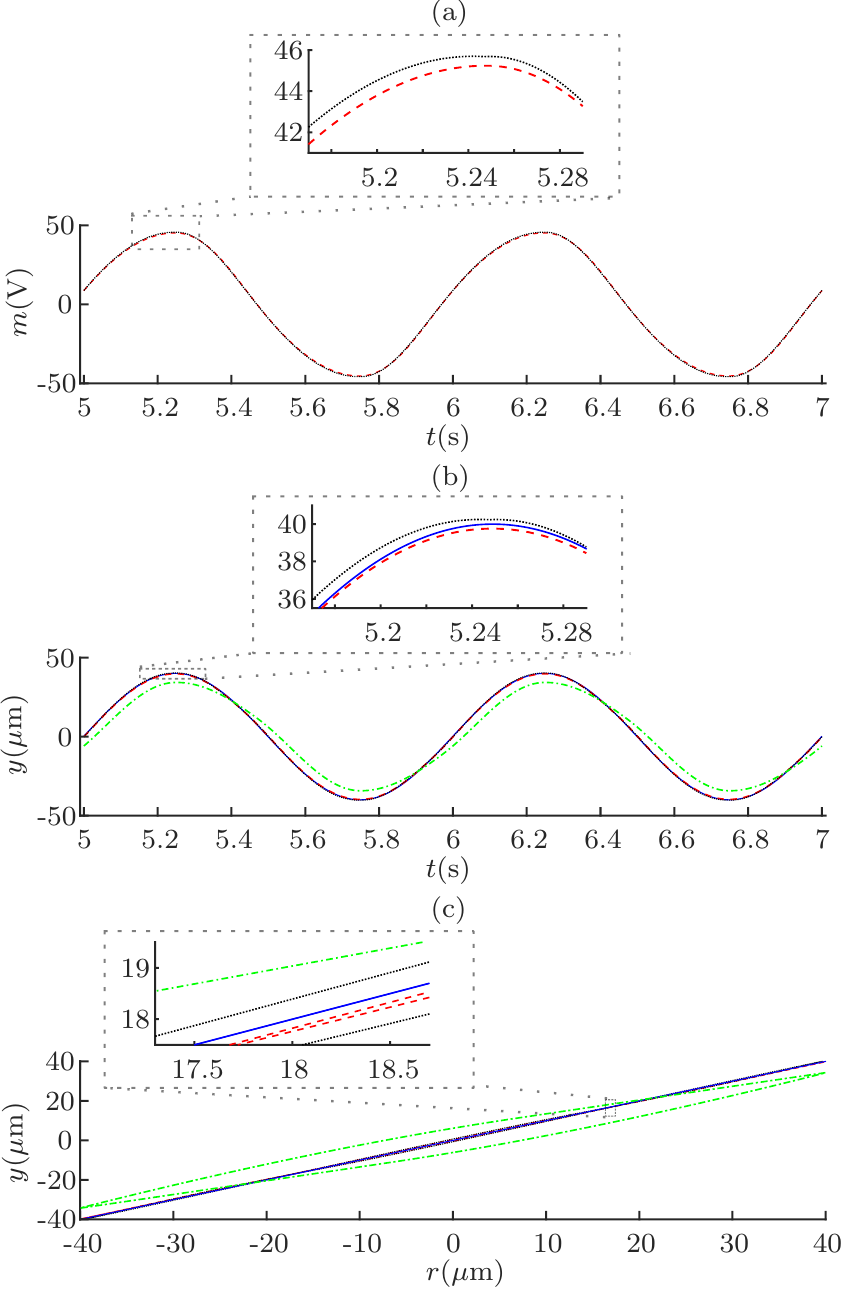} \vspace{-2mm}
		\caption{Hysteresis compensation for the piezoelectric actuator (\ref{Eq:NumericalResults:System}). (a)~Compensation inputs, (b)~temporal responses and in (c)~hysteresis loops. 
			\hbox{(\textcolor{red}{- -})}~results obtained with  compensator (\ref{Eq:NumericalResults:InputCompensation:Cms}) 
			(\textcolor{black}{$\cdots$})~results with compensator (\ref{Eq:NumericalResults:InputCompensation:Cci}),
			(\textcolor{green}{-$\,\cdot\,$-})~system output without compensation, and (\textcolor{blue}{---})~displacement 
			reference $r=40\sin (2\pi t)\,\mu$m.} % thick lines.
		\label{Fig:NumericalResults:Compensated_System_For_C_ms_C_ci}
	}
\end{figure}

The results are summarized in figure\,\ref{Fig:NumericalResults:Compensated_System_For_C_ms_C_ci}(b).
From the hysteresis loops shown in figure\,\ref{Fig:NumericalResults:Compensated_System_For_C_ms_C_ci}(c),
it is clear that the compensators enforced a practically linear relation between the reference and the
output. This would greatly facilitate the design and increase the performance of a feedback controller.

The accuracy achieved by each compensator was quantified by the {\rm MAPE} index (\ref{Eq:MAPE}). 
In order to quantify the compensation effort, the normalized sum of the absolute variation
of the input ({\rm NSAVI})
\begin{equation}
\label{Eq:Compensation_Effort}
%{\rm IAVI}=\int_{0}^{t_f}  \bigg|\frac{dm}{dt}\bigg|dt. \label{Eq:Compensation_Effort}
{\rm NSAVI}=\sum_{k=1}^{N-1} \frac{\big|m_{k+1}-m_k\big|} {\big|r_{k+1}-r_k\big|},
\end{equation}

\noindent
is calculated. These indices are shown in Table\,\ref{Tabel:NumericalResults:Compensation_System:PerformanceIndex}.

\begin{table}[htb]
	\centering
	\caption{Performance of the compensation step. Simulation results.} % Performance of the compensators designed for a bench test.
	\label{Tabel:NumericalResults:Compensation_System:PerformanceIndex} \setlength\tabcolsep{3.8pt} % default value: 6pt
	\begin{tabular}{c|c|ccc}
		%\hline
		\textbf{Design Strategy }&				\textbf{Compensator}		                       & {\rm MAPE} & {\rm NSAVI} \\ \hline 
		Section\,\ref{Subsection:Compensator:Cms} & (\ref{Eq:NumericalResults:InputCompensation:Cms}) & $0.322$    & $1.13$	  \\ \hline
		Section\,\ref{Subsection:Compensator:Cci} & (\ref{Eq:NumericalResults:InputCompensation:Cci}) & $0.425$	& $1.14$      \\  \hline
		Black-box							   &  not shown										   & $1.819$    & $1.15$      \\ \hline
		\multicolumn{2}{c|}{no compensation}			   							   & $6.536$    & $1.00$      \\		
	\end{tabular}
	% $9059.4$, $9123.5$, $7999.7$
\end{table}

The results shown in figure\,\ref{Fig:NumericalResults:Compensated_System_For_C_ms_C_ci} and
Table\,\ref{Tabel:NumericalResults:Compensation_System:PerformanceIndex} indicate that
the compensators may provide a significant improvement in the tracking performance of system
(\ref{Eq:NumericalResults:System}). The tracking error was reduced by about $93\%$ at
the cost of a $14\%$ increase in the compensation effort. Although the compensator strategies
yield similar results, the design strategy of section\,\ref{Subsection:Compensator:Cms}
yielded results with lower compensation effort and tracking error.

\begin{figure}[htb]%[htb]
	\centering{
		\includegraphics[scale=0.97]{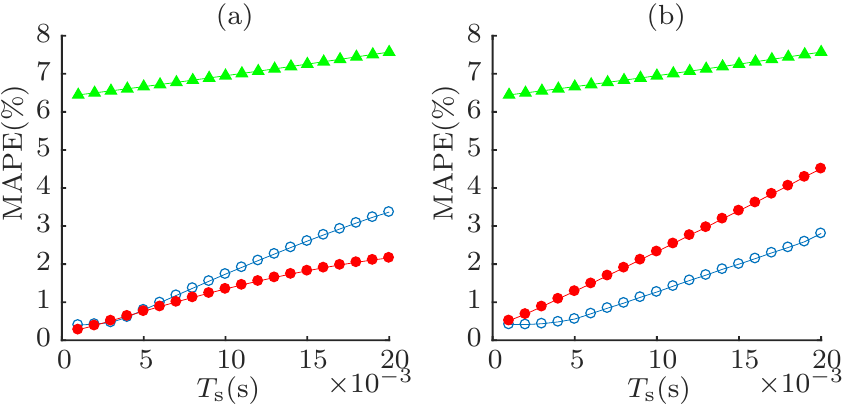} \vspace{-2mm}
		\caption{MAPE index (\ref{Eq:MAPE}) computed for the models and compensators described, respectively, by equations (a)~(\ref{Eq:NumericalResults:IdentifiedModel:Cms}) and (\ref{Eq:NumericalResults:InputCompensation:Cms}); (b)~(\ref{Eq:NumericalResults:IdentifiedModel:Cci}) and (\ref{Eq:NumericalResults:InputCompensation:Cci}). (\textcolor{blue}{$\circ$})~model and (\textcolor{red}{$\bullet$})~tracking accuracies. 
			(\textcolor{green}{$\blacktriangle$})~accuracy of uncompensated system.} % thick lines.
		\label{Fig:NumericalResults:Analysis_Sampling_Time}
	}
\end{figure}

To further characterize the performance of the proposed designs, the influence of the
sampling time $T_{\rm s}$ is also investigated. In figure~\ref{Fig:NumericalResults:Analysis_Sampling_Time},
it can be seen that the model accuracy somewhat deteriorates as $T_{\rm s}$ is increased.
It should be noted that even the largest values of $T_{\rm s}$ in
figure~\ref{Fig:NumericalResults:Analysis_Sampling_Time} are still comfortably small in terms
of the sampling theorem. However, since one of the regressors is the first difference of the input,
then the identification of systems with hysteresis seems to be particularly sensitive to the sampling
time \cite{L_Junior_etal2017}. Another conclusion that can be drawn from figure~\ref{Fig:NumericalResults:Analysis_Sampling_Time}
is that, for both design strategies, the compensation performance is correlated to the model accuracy,
and that the strategy in section\,\ref{Subsection:Compensator:Cms} (figure\,\ref{Fig:NumericalResults:Analysis_Sampling_Time}(a)) 
is somewhat less sensitive to such accuracy.

\begin{figure}[htb]%[htb]
	\centering{
		\includegraphics[scale=1]{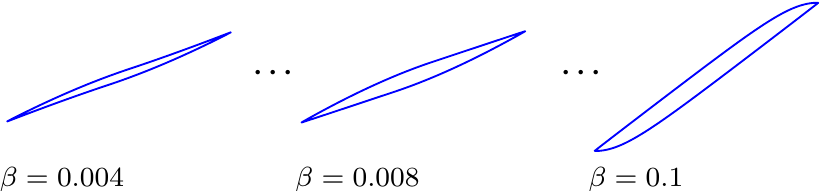} \vspace{-2mm}
		\caption{Bouc-Wen hysteresis loops within the investigated range.} % thick lines.
		\label{Fig:NumericalResults:GeometricShapes}
	}
\end{figure}

Finally, the same analysis was carried out for situations with different shapes of the hysteresis
loop varying $\beta$ in the range $0.004 \le \beta \le 0.1$ with increments of $\Delta=0.002$
(see figure\,\ref{Fig:NumericalResults:GeometricShapes}). The results are quite similar to the ones
described so far and are not shown.

%==================================================================
\section{Experimental Results}
\label{Section:ExperimentalResults}

Both identification and compensation strategies are now applied to an experimental pneumatic control valve.
This type of actuator is widely used in industrial processes, for which control performance can degrade
significantly due to valve problems caused by nonlinearities \cite{Srinivasan_Rengaswamy2005} such as
friction \cite{Romano_Garcia2011,Baeza_Garcia2018}, dead-zone, dead-band and hysteresis \cite{Choudhury_etal2008}.
Hence, in this section we aim at compensation hysteresis using the developed techniques.

The measured output is the stem position of the pneumatic valve and the input is a signal that, after passing
V/I and I/P conversion, becomes a pressure signal applied to the valve. The sampling time is $T_{\rm s}=0.01\,{\rm s}$.
For model identification, the input is set as white noise low-pass filtered at $0.1\,{\rm Hz}$. For model validation,
the input is a sinusoid with frequency $0.1\,{\rm Hz}$. Both data sets are $200\,{\rm s}$ long ($N=20000$). The
identification of the direct $\mathcal{M}$ and inverse $\breve{\mathcal{M}}$ models was performed as in section\,\ref{Section:NumericalResults}. The pool of candidate terms is generated with $\ell=3$, $n_{y}=1$ and $n_{u}=2$.
The model parameters are estimated using (\ref{Eq:ConstrainedParameters_Estimator}) in order to comply with
Lemma~\ref{D1}.

\begin{figure}[htb]%[htb]
	\centering{
		\includegraphics[scale=0.97]{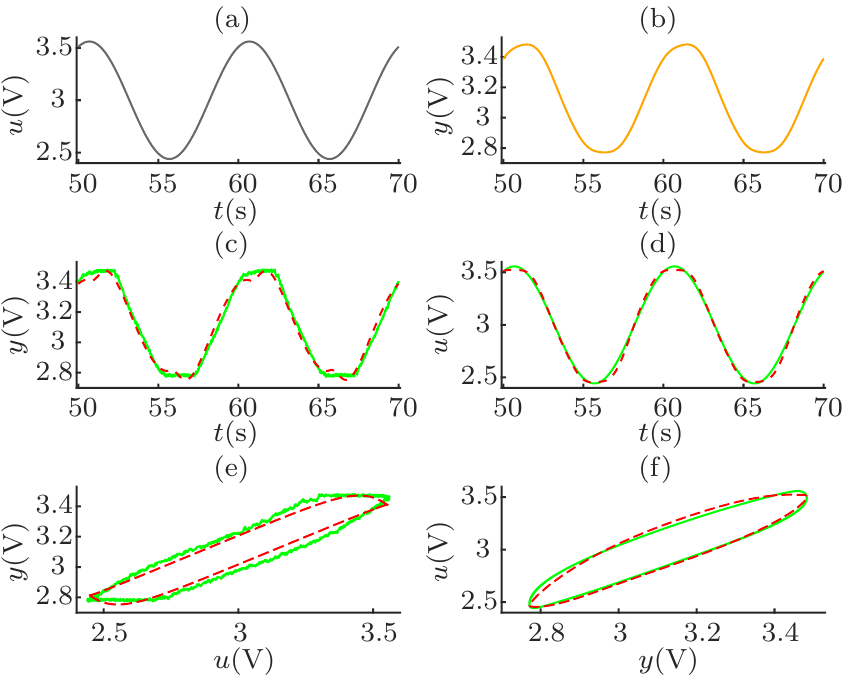} \vspace{-2mm}
		\caption{Left column refers to model (\ref{Eq:ExperimentalResults:IdentifiedModel:Cms}) and right column to model (\ref{Eq:ExperimentalResults:IdentifiedModel:Cci}). (a)~input $u_k{=}0.56\sin(0.2\pi k)+3\,{\rm V}$ and 
			(c)~the corresponding measured output (\textcolor{green}{---}) $y$ and (\textcolor{red}{- -}) model
			(\ref{Eq:ExperimentalResults:IdentifiedModel:Cms}) free-run simulation; (b)~smoothed version of $y$ in (c);
			(d)~the corresponding output which is $u_k$ in (a) and (\textcolor{red}{- -}) model
			(\ref{Eq:ExperimentalResults:IdentifiedModel:Cci}) free-run simulation. (e) and (f) show the same data as (c) and (d),
			respectively.} % thick lines.
		\label{Fig:ExperimentalResults:Valid_Models}
	}
\end{figure}

The estimated model $\mathcal{M}$ is
\begin{eqnarray}
\label{Eq:ExperimentalResults:IdentifiedModel:Cms}
y_k&=y_{k-1}-19.76\phi_{1,\,k-2}+19.32\phi_{1,\,k-1}\nonumber\\
&{+}9.44\phi_{2,\,k{-}2}\phi_{1,\,k{-}2}u_{k{-}2}{-}12.61\phi_{2,\,k{-}2}\phi_{1,\,k{-}2}y_{k{-}1}, 
\end{eqnarray}

\noindent
and the inverse model $\breve{\mathcal{M}}$ is
\begin{eqnarray}
\label{Eq:ExperimentalResults:IdentifiedModel:Cci}
\hat{u}_k&=\hat{u}_{k{-}1}+86.67\breve{\phi}_{1,\,k{-}1}-85.02\breve{\phi}_{1,\,k{-}2}{-}0.98\breve{\phi}_{1,\,k{-}1}y_{k{-}2} \nonumber\\
&+1.72\breve{\phi}_{2,\,k{-}2}\breve{\phi}_{1,\,k{-}2}y_{k{-}2}{-}1.13\breve{\phi}_{2,\,k{-}2}\breve{\phi}_{1,\,k{-}2}\hat{u}_{k{-}1},
\end{eqnarray}

\noindent
which was estimated from a smoothed version of $y_k$ obtained by quadratic regression. This
is done only to estimate $\breve{\mathcal{M}}$ to avoid the error-in-the-variables problem,
since $y_k$ serves as the input for $\breve{\mathcal{M}}$. Each model performance is given
in figure\,\ref{Fig:ExperimentalResults:Valid_Models} and Table\,\ref{Tabel:ExperimentalResults:IdentifiedModels:PerformanceIndex}.

\begin{table}[htb]%[htb]
	\centering
	\caption{Performance of the modeling step. Experimental results.} %Performance index of the identified models for a bench test.
	\label{Tabel:ExperimentalResults:IdentifiedModels:PerformanceIndex} %\setlength\tabcolsep{6.0pt} % default value: 6pt
	\begin{tabular}{c|c|c}
		%\hline
		\textbf{Design Strategy }&\textbf{Model}&{\rm MAPE} \\ \hline 
		Section\,\ref{Subsection:Compensator:Cms} & (\ref{Eq:ExperimentalResults:IdentifiedModel:Cms}) & $3.926$ \\ \hline
		Section\,\ref{Subsection:Compensator:Cci} & (\ref{Eq:ExperimentalResults:IdentifiedModel:Cci}) & $2.374$ \\
	\end{tabular}
\end{table}

Models (\ref{Eq:ExperimentalResults:IdentifiedModel:Cms}) and (\ref{Eq:ExperimentalResults:IdentifiedModel:Cci})
are used to implement the strategies described in sections~\ref{Subsection:Compensator:Cms} and \ref{Subsection:Compensator:Cci},
thus yielding, respectively, the compensation inputs
\begin{eqnarray}
\label{Eq:ExperimentalResults:InputCompensation:Cms}
%\hspace{-2.0mm}
m_k{=}\frac{1}{19.32}&\!\Big[r_{k{+}1}{-}r_{k}{+}19.32m_{k{-}1}{+}19.76[m_{k{-}1}{-}m_{k{-}2}] \nonumber\\
&-9.44{\rm sign}(m_{k-1}-m_{k{-}2})[m_{k{-}1}-m_{k{-}2}]m_{k{-}1} \nonumber\\
&{+}12.61{\rm sign}(m_{k{-}1}{-}m_{k{-}2})[m_{k{-}1}{-}m_{k{-}2}]r_{k}\Big]\!,
\end{eqnarray}

\noindent
and 
\begin{eqnarray}
\label{Eq:ExperimentalResults:InputCompensation:Cci}
\breve{m}_k&=\breve{m}_{k{-}1}{+}86.67[r_{k{+}1}-r_{k}]{-}85.02[r_{k}-r_{k{-}1}] \nonumber\\
&-0.98[r_{k{+}1}-r_{k}]r_{k}+1.72{\rm sign}(r_{k}-r_{k{-}1})[r_{k}-r_{k{-}1}]r_{k}\nonumber\\
&-1.13{\rm sign}(r_{k}-r_{k{-}1})[r_{k}-r_{k{-}1}]\breve{m}_{k-1}. 
\end{eqnarray}

\begin{figure}[htb]%[htb]
	\centering{
		\includegraphics[scale=0.97]{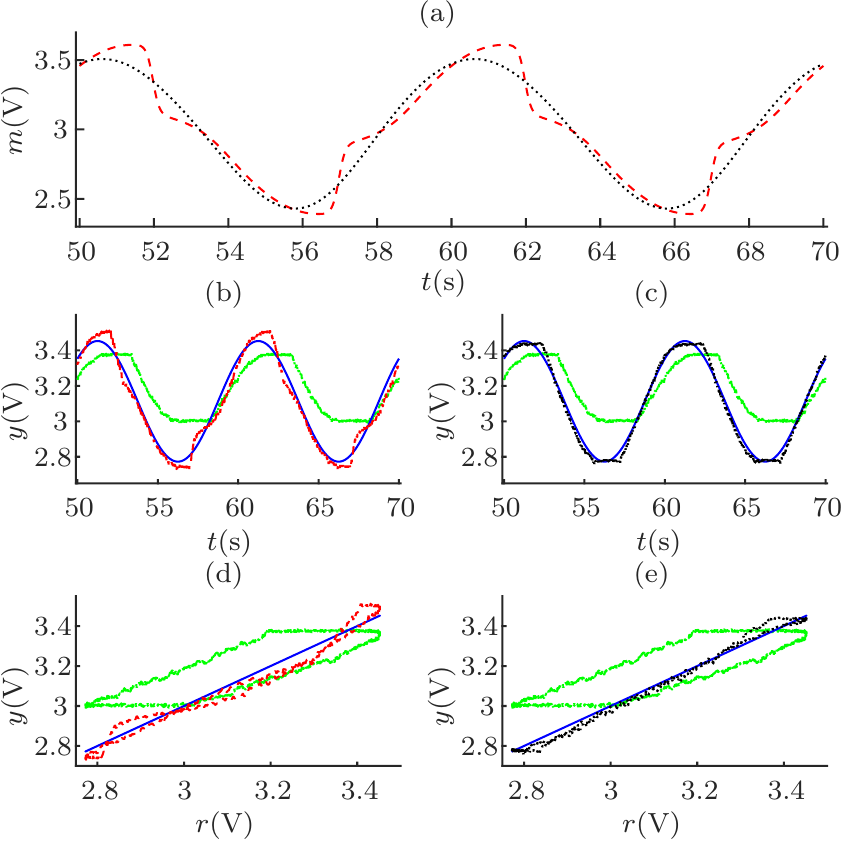} \vspace{-2mm}
		\caption{Hysteresis compensation for the pneumatic valve. (a)~Compensation inputs, (b) and (c)~its temporal responses and in (d) and (e)~ the hysteresis loops. 
			\hbox{(\textcolor{red}{- -})}~illustrates the results obtained with compensator (\ref{Eq:ExperimentalResults:InputCompensation:Cms}),
			(\textcolor{black}{$\cdots$})~refers to the results by using compensator (\ref{Eq:ExperimentalResults:InputCompensation:Cci}),
			(\textcolor{green}{-$\,\cdot\,$-})~the system output without compensation, and (\textcolor{blue}{---})~the reference 
			$r{=}0.56\sin (0.2\pi t){+}3\,{\rm V}$.} % thick lines.
		\label{Fig:ExperimentalResults:Compensated_Valve}
	}
\end{figure}

The results of the experiment are shown in figure\,\ref{Fig:ExperimentalResults:Compensated_Valve}
and assessed in Table\,\ref{Tabel:ExperimentalResults:Compensation_System:PerformanceIndex}. Note that
both approaches significantly reduce the tracking error. Based on both numerical and experimental
results, it seems that the performance of the compensators is directly related to the accuracy of the
identified model; see Table\,\ref{Tabel:ExperimentalResults:IdentifiedModels:PerformanceIndex} and Table\,\ref{Tabel:ExperimentalResults:Compensation_System:PerformanceIndex}. Hence, as before, these results
suggest that the compensation effort tends to be lower and more effective whenever the identified models
are more accurate.

The compensation produced by (\ref{Eq:ExperimentalResults:InputCompensation:Cci}) is smoother than the
one obtained with (\ref{Eq:ExperimentalResults:InputCompensation:Cms});
see figure\,\ref{Fig:ExperimentalResults:Compensated_Valve}(a). This occurs because, for the compensator (\ref{Eq:ExperimentalResults:InputCompensation:Cci}), the argument of the sign function depends on the
difference of the reference signal, while, for the compensator (\ref{Eq:ExperimentalResults:InputCompensation:Cms}),
it depends on the difference of the autoregressive variable which usually produces stronger oscillations and sudden
changes; see figure\,\ref{Fig:ExperimentalResults:Compensated_Valve}(a), e.g. in the range of $51{-}53\,{\rm s}$.
As a result, larger compensation effort is required as quantified by NSAVI (\ref{Eq:Compensation_Effort}) in Table\,\ref{Tabel:ExperimentalResults:Compensation_System:PerformanceIndex}.

\begin{table}[htb]
	\centering
	\caption{Performance of the compensation step. Experimental results.} % Performance of the compensators designed for a bench test.
	\label{Tabel:ExperimentalResults:Compensation_System:PerformanceIndex} \setlength\tabcolsep{3.8pt} % default value: 6pt
	\begin{tabular}{c|c|ccc}
		%\hline
		%\br
		\textbf{Design Strategy }&				\textbf{Compensator}		     &{\rm MAPE}      & {\rm NSAVI} \\ \hline 
		Section\,\ref{Subsection:Compensator:Cms} & (\ref{Eq:ExperimentalResults:InputCompensation:Cms}) & $5.514$	& $1.81$	\\ \hline
		Section\,\ref{Subsection:Compensator:Cci} & (\ref{Eq:ExperimentalResults:InputCompensation:Cci}) & $2.939$ & $1.61$    \\  \hline
		\multicolumn{2}{c|}{no compensation}			   	 										  & $18.602$& $1.00$    \\		
	\end{tabular}
	% $9059.4$, $9123.5$, $7999.7$
\end{table}

%==================================================================
\section{Conclusions}
\label{Section:Conclusions}

This work addressed the problems of identification and compensation of
hysteretic systems. In the context of {\it system identification}, the
contribution is twofold. First, we build models with regressors that use
the sign function of the first difference of the input, as proposed by
\cite{Martins_Aguirre2016}, and present an additional condition in order
to guarantee a {\it continuum of equilibrium points} at steady-state,
which is an important ingredient for hysteresis \cite{Bernstein2007,Morris2011}.
To this aim, a particular constraint on the parameters is presented in Lemma~\ref{P1}.
As a consequence, the identified models are able to describe both dynamical and static
features of the hysteresis nonlinearity. Second, following a {\it quasi-static analysis}
of these models, a schematic framework is proposed to explain how the hysteresis loop
occurs on the input-output plane (see Figure~\ref{F2}).

In the context of {\it hysteresis compensation}, this paper introduces two
strategies to design compensators. An important aspect of such procedures is that
they show how to restrict the pool of candidate regressors aiming at solving the
compensation problem. Such strategies are not limited to hysteresis and can be
extended to other nonlinearities.

The effectiveness of the compensation schemes is illustrated by means of numerical and
experimental tests. For the strategy described in Section\,\ref{Subsection:Compensator:Cms},
the compensation law is obtained from the identified model by simple algebraic manipulations.
In the case of the strategy introduced in Section\,\ref{Subsection:Compensator:Cci}, the
compensators are identified directly from the data. This, however, leads to models that are
noncausal by nature. Guidelines to circumvent this problem are provided. The compensators
designed by both strategies can be readily employed in online compensation schemes.

As a general remark, we observed in all our examples that the quality of the achieved compensation is 
correlated with the accuracy of the identified model
(compare Table\,\ref{Tabel:NumericalResults:IdentifiedModels:PerformanceIndex}
with Table\,\ref{Tabel:NumericalResults:Compensation_System:PerformanceIndex} and
Table\,\ref{Tabel:ExperimentalResults:IdentifiedModels:PerformanceIndex} with
Table\,\ref{Tabel:ExperimentalResults:Compensation_System:PerformanceIndex}). 
Finally, we noticed that the identified models have a discontinuity due to the
sign function used in some regressors. When the model
has many such terms, it sometimes happens that the compensation signal presents abrupt transitions.
The use of smoother functions in place of the sign function, in order to alleviate this problem, will be investigated
in the future.

%==================================================================
\section*{Acknowledgments}

The authors would like to thank Arthur N Montanari for the insightful discussions.
PEOGBA, BOST and LAA gratefully acknowledge financial support from CNPq
(Grants Nos. 142194/2017-4, 310848/2017-2 and 302079/2011-4) and 
FAPEMIG (TEC-1217/98).

%==================================================================
\section*{References}

\bibliography{references}

\end{document}